\input amstex
\documentstyle{amsppt}
\magnification=1200


\NoRunningHeads \TagsOnRight \noindent
\medskip
\topmatter
\title
ON SCATTERING BY A CYLINDRICAL TRAP IN CRITICAL CASE
\endtitle
\thanks
Supported by RFBR grants 02-01-00768, 00-15-96038 and by MERF
grant Å00-1.0-53.
\endthanks
\author
Rustem R.~Gadyl'shin
\endauthor

 \abstract We consider a two-dimensional analogue of Helmholtz resonator with walls of
finite thick\-ness in the critical case when there exists an
eigenfrequency equalling to the limit of poles generated by both
the bounded component of the resonator and by the narrow
connecting channel. Under assumption that the limit
eigenfrequency is simple one of the bounded component,
asymptotics of two poles converging to this eigenfrequency are
constructed by using the method of matching asymptotic
expansions.  The explicit formulas for the leading terms of
asymptotics for poles and for the solution of the scattering
problem are obtained.

\endabstract
\endtopmatter

\document

We study the case where the cross-section of the cylindrical
scattering object is asym\-p\-to\-ti\-cal\-ly homeomorphic to
ring. Perturbed domain is formed by both the components of the
limiting cross-section exterior (the bounded domain $\Omega^{in}$
and unbounded domain  $\Omega^{ex}$) and the narrow channel
$\varkappa_\varepsilon$ connecting them and having the "diameter"
of order  $\varepsilon<<1$. The corresponding mathematical model
(both for perturbed and limits problem) is described by the
Neumann boundary value problem for the Helmholtz equation. It is
also known ([1]) that the analytic continuation of the perturbed
solution in this case (in contrast to the case of the Dirichlet
boundary condition ([2]))  has two series of poles with small
imaginary parts. The limiting set for the first series is a set
$\Sigma^{in}$ of eigenfrequencies (square roots of  the
eigenvalues) of the Neumann boundary value problem for  $-\Delta$
in $\Omega^{in}$ (the limit internal problem). The limiting set
for the second series  is $\Sigma^{ch}=\{m\pi/h\}_{m=1}^\infty$,
$h$ is the length of the connecting channel. Brown, Hislop and
Martinez [3] considered two situations assuming that a limiting
frequency $k_0$ is a simple eigenfrequency belonging to
$\Sigma^{in}$ and separated from $\Sigma^{ch}$ or, on the
contrary, $k_0\in\Sigma^{ch}\backslash\Sigma^{in}$. For these
situations they showed that the analytic continuation of the Green
function for the perturbed problem has the only simple pole
converging to $k_0$ as $\varepsilon\to0$ and there is only one
generalized eigenfunction associated with this pole. In [4] the
explicit formulae for the leading terms of both these
characteristics and  peaks for solution to the scattering problem
for {\it real frequencies} $k$ close to $k_0$ were obtained
employing the method of matched asymptotic expansions [5--7].
These formulae implied that for both cases the solution of the
perturbed  problem differ from one of the limiting problem at a
quantity $O(1)$ in the resonator exterior (i.e. outside bounded
component and connecting channel). Inside the bounded component
(trap) the behavior of the perturbed solution was really different
for these cases, for frequencies close to
$k_0\in\Sigma^{in}\backslash\Sigma^{ch}$ it was of order
$O(\varepsilon^{-1})$, while for frequencies close to
$\Sigma^{ch}\backslash\Sigma^{in}$ it was bounded.

In this paper we analyze a critical case assuming that the
limiting frequency $k_0$ is a simple eigenfrequency of interior
limiting problem and, at the same time, belong to $\Sigma^{ch}$.
The results of this work were announced   in [8].

\head \S1. Description of the problem, preliminary notes and
formulation of the results
\endhead

Let $\Omega^{in}$ and $\Omega$ be simply connected bounded
domains in ${\Bbb R}^2$, $\overline\Omega^{in}\subset\Omega$,
$\Omega^{ex}={\Bbb R}^{2}\backslash\overline\Omega$,
$\partial\Omega^{in(ex)}\in C^\infty$, $x=(x_1,x_2)$. The
domains $\Omega^{in}$ and $\Omega^{ex}$ are supposed to coincide
with the half-plane $x_2>0$ in the neighborhood of the origin
and the half-plane $x_2<-h$ in the neighborhood of a point
$x^{0}=(0,-h)$, respectively. We postulate that  the interval
$(-h,0)$ lying on the axis $Ox_2$  not to contain the points
from $\Omega^{in}\cup\Omega^{ex}$. The domains $\Omega ^{in}$
and $\Omega^{ex}$ are the interior and exterior of the resonator
$$
\Omega _\varepsilon =\Omega^{in}\cup\Omega^{ex}
\cup\varkappa_\varepsilon,
$$
respectively, where $\varkappa _\varepsilon =
(\varepsilon\omega_-,\varepsilon\omega_+)\times[-h,0]$ is the
connecting channel, $\omega_-<\omega_+$ are arbitrary constants.

It is known  that the scattering of both the  $H$-polarized
fields on an ideal conductive cylinder with cross-section
$\Omega_\varepsilon$, and the plane acoustic waves on an ideal
rigid cylinder with cross-section $\Omega_\varepsilon$ are
described by the solution of a boundary value problem

$$ \align (\Delta+k^2)u_\varepsilon & =F,\quad
x\in\Omega_\varepsilon,\qquad {\partial
u_\varepsilon\over\partial\nu} = 0,\quad
x\in\partial\Omega_\varepsilon,\tag1.1\\ u_\varepsilon &
=O\left({1\over r^{1/2}}\right),\quad{\partial
u_\varepsilon\over\partial r} - iku_\varepsilon=o\left({1\over
r^{1/2}}\right),\quad r\rightarrow\infty,\tag1.2
\endalign
$$
where  $r=\vert x\vert$, $\nu$  is the outward normal, $k>0$,
and $F$ is a square integrable function with finite support in
$\Omega^{ex}$. For the acoustic scattering, $u_\varepsilon$ is
the potential of the velocity. For $H$-polarization  $F$ denotes
the third component of the vector $-\hbox{rot{\bf j}}$ in the
case where the current vector $\hbox{\bf j}$ is perpendicular to
a generatrix and electromagnetic field  is of the form
$\hbox{{\bf H}}_\varepsilon=\{0,0,u_\varepsilon\}$, $\hbox{\bf
E}_\varepsilon=-ik^{-1}\left(\hbox{rot\,{\bf
H}}_\varepsilon-\hbox{\bf j}\right)$. Throughout  in what
follows by interior (exterior) limiting problem we mean Neumann
value problem for the Helmholtz equation in $\Omega^{in}$ (in
$\Omega^{ex})$.

It is known (see, for instance, [9]) that for positive $k$ the
boundary value problem  (1.1),  (1.2) and the exterior limiting
problem are unique solvable, and their Green functions admit
analytic continuations in a complex plane with a cut along the
negative real semi-axis, which, for fixed $\varepsilon$, have
discrete sets of poles $\Sigma_\varepsilon$ and $\Sigma^{ex}$,
respectively, lying below the real axis. As it was mentioned
above, it was proved in [1] that in each small neighborhood of a
nonzero element
$k_0\in\Sigma=\Sigma^{in}\cup\Sigma^{ex}\cup\Sigma^{ch}$, there
exists pole $\tau_\varepsilon\in\Sigma_\varepsilon$, for
$\varepsilon$ enough small,  and, visa versa, if a compact $K$
is separated from $\Sigma$, then
$K\cap\Sigma_\varepsilon=\emptyset$, for all $\varepsilon$
enough small.

In cases, when $k_0\in\Sigma^{in}_1\backslash\Sigma^{ch}$,
$\Sigma^{in}_1$ is the set of simple nonzero eigenfrequencies of
the interior limiting problem, and when
$k_0\in\Sigma^{ch}\backslash\Sigma^{in}$,  there exists one pole
(in each case) converging to  $k_0$ ([4]).  Since we consider
solutions of (1.1), (1.2) for $k>0$, and the pole
$\tau_\varepsilon$ is complex, in both cases the solutions are
most perturbed when
$
k=k(\varepsilon)=\hbox{\rm Re}\,\tau_\varepsilon+O(\hbox{\rm
Im}\,\tau_\varepsilon).
$
We will call such positive frequencies as the peak regime. Let
$S(t)$ and $S^{ex}(t)$ be the disks of radius  $t$ with their
centers at the origin  an at  $x^{(0)}$, respectively,
$G^{in}(x,y,k)$ ($G^{ex}(x,y,k)$) be the Green' function of the
interior (exterior)  limiting problem, $u^{ex}(x;k)$ be the
solution of the exterior  limiting problem, and $\psi$ be the
eigenfunction of the interior  limiting problem, associated with
a simple eigenfrequency $k_0\in\Sigma_1^{in}$ and  normalized in
$L_2(\Omega^{in})$. In [4] it was shown that at the peak regime
the leading terms of asymptotics of the perturbed boundary value
problem  (1.1), (1.2) reads as follows:
$$ \align u_\varepsilon(x;k(\varepsilon)) &
\sim{1\over\varepsilon}A_1 \psi(x),\quad x\in\Omega^{in}\backslash
S(\varepsilon^{1/2}),\tag1.3\\ u_\varepsilon(x;k(\varepsilon)) &
\sim A_2 G^{ex}(x,x^{(0)},k_0)+u^{ex}(x;k_0),\quad
x\in\left(\Omega^{ex}\backslash
S^{ex}(\varepsilon^{1/2})\right)\cap S(R)\tag1.4
\endalign
$$ for $k(\varepsilon)\to
k_0\in\Sigma_1^{in}\backslash\Sigma^{ch}$ and $$ \align
u_\varepsilon(x;k(\varepsilon)) & \sim B_1\,G^{in}(x,0,k_0),\qquad
x\in\Omega^{in} \backslash S(\varepsilon^{1/2}),\tag1.5\\
u_\varepsilon(x;k(\varepsilon)) & \sim B_2\,G^{ex}(x,x^{(0)},k_0)
+ u^{ex}(x,k_0),\quad x\in\left(\Omega^{ex}\backslash
S^{ex}(\varepsilon^{1/2})\right)\cap S(R)\tag1.6
\endalign
$$ for $k(\varepsilon)\to k_0\in\Sigma^{ch}\backslash\Sigma^{in}$,
where $A_j$ and $B_j$ are some constants calculated  explicitly
and $R>0$ is an arbitrary number.

From (1.4) and (1.6) it follows that at peak regimes in both
cases the solutions of perturbed problem differs from the
solution of the exterior limiting problem at $O(1)$  in
$\Omega^{ex}$. Exactly this difference was observed by Rayleigh
for classical Helmholtz resonator (which is a sphere with the
small connecting opening) in [10]. We call this effect as the
exterior resonance. On the other hand, it follows from (1.3) and
(1.5) that for peak frequencies the solutions to perturbed
problem distinguish essentially each from other in
$\Omega^{in}$. For $k(\varepsilon)\to
k_0\in\Sigma^{ch}\backslash\Sigma^{in}$, the solution is bounded
and, for $k(\varepsilon)\to
k_0\in\Sigma_1^{in}\backslash\Sigma^{ch}$, it increases as
$O(\varepsilon^{-1})$. We call the effect (1.3) as the interior
resonance. Note, for the three-dimensional resonator with walls
of finite thickness, in [11] the resonance is mean  exactly in
this sense, and the asymptotics of solutions, for
$k(\varepsilon)\to k_0\in\Sigma^{ch}\backslash\Sigma^{in}$ and
$k(\varepsilon)\to k_0\in\Sigma_1^{in}\backslash\Sigma^{ch}$,
were constructed in [12], [13]. Since the difference between
(1.3) and (1.5) is great enough, it is interesting to consider
the case  $k_0\in\Sigma^{in}_1\cap\Sigma^{ch}$. Note that this
critical case suggested to be analyzed is not very degenerated,
because, for  the fixed "main"  trap $\Omega^{in}$, such
situation can be easily achieved by a corresponding change for
the length of the connecting channel (see definition of
$\Sigma^{ch}$).

Hereafter we will employ the notations: $\omega$ is an interval
$(\omega_-,\omega_+)$, $|\omega|=\omega_+-\omega_-$,
$x^*=(x_1,-x_2)$,
$$\sigma=\lim_{R\to\infty}\int\limits_{\partial
S(R)}|G^{ex}(x,x^{(0)},k_0)|^2\,ds.
$$

The main goal of this work is to prove the following statement.

\proclaim{Theorem 1.1} Let $k_0={m\pi\over h}\in\Sigma^{in}_1\cap
\Sigma^{ch}$. Then

a) there exist two poles
$\tau_\varepsilon^{(n)}\in\Sigma_\varepsilon$ ($n=1,2$),
converging to $k_0$ and having  asymptotics:
$$
\tau_\varepsilon^{(n)}=k_0+\sum_{i=1}^\infty\sum_{j=0}^{i-1}
\varepsilon^{i/2}\ln^j\varepsilon\,\tau_{i,j}^{(n)},\tag1.7 $$ $$
\aligned &
\tau_{1,0}^{(n)}=(-1)^n\psi(0)\left({|\omega|\over2h}\right)^{1/2},\qquad
\tau_{2,1}^{(n)}= (-1)^n{4k_0\over\pi\psi(0)}
\left({|\omega|\over2h}\right)^{1/2},\\ & \hbox{\rm
Im}\,\tau_{2,0}^{(n)}=-{1\over2} {|\omega|\over h}k^2_0\sigma ;
\endaligned\tag1.8
$$

b) for $k$ close to  $k_0$ the solution of  (1.1), (1.2) and its
analytic continuation can be represented as:
$$ u_\varepsilon(x;k)=-\sum_{n=1}^2{\Psi_\varepsilon^{(n)}(x)\over
\left(\tau_\varepsilon^{(n)}\right)^2-k^2} \int_{\Bbb R^2}
\Psi_\varepsilon^{(n)}(y)F(y)\,dy+\tilde u_\varepsilon(x;k)\tag1.9
$$ where $\tilde u_\varepsilon$ is a holomorphic on $k$ function.
If $\hbox{\rm supp} F\subset\Omega^{ex}$, then  in $\Omega^{ex}$
$\tilde u_\varepsilon$ converges to the solution of the exterior
limiting problem in $L_{2,loc}(\Omega^{ex})$ (i.e., on any
compact set from $\Omega^{ex}$) and it converge to zero in
$\Omega^{in}\cup\varkappa_\varepsilon$ in $L_2$ norm;

c) for $\Psi_\varepsilon^{(n)}$ the following asymptotics hold
in $L_{2,loc}({\Bbb R}^2)$ $$ \allowdisplaybreaks \align
&\Psi_\varepsilon^{(n)}(x)
={(-1)^n\over\sqrt{2}}\psi(x)+o(1),\qquad
x\in\Omega^{in}\backslash S(\varepsilon^{1/2}),\\
&\Psi_\varepsilon^{(n)}(x)
={(-1)^n\over\sqrt{2}}\psi(0)(1+o(1)),\qquad x\in
S(2\varepsilon^{1/2}),\\ &\Psi_\varepsilon^{(n)}(x)
={1\over\varepsilon^{1/2}}\left({1\over
h|\omega|}\right)^{1/2}(\sin(k_0x_2) +o(1)),\qquad
x\in\varkappa_\varepsilon\backslash (S^{ex}(\varepsilon^{1/2})
\cup S(\varepsilon^{1/2})),\\ &\Psi_\varepsilon^{(n)}(x)
=\varepsilon^{1/2} {(-1)^{m+1}k_0\over\pi}\left({|\omega|\over
h}\right)^{1/2}\left(\ln\varepsilon+
{\pi\over|\omega|}X\left({\left(x-x^{(0)}\right)^*\over\varepsilon}
\right)+o(1)\right), \\&\qquad\qquad\qquad x\in
S^{ex}(2\varepsilon^{1/2}),
\\
&\Psi_\varepsilon^{(n)}(x)
=\varepsilon^{1/2}\left((-1)^mk_0\left({|\omega|\over
h}\right)^{1/2} G^{ex}(x,x^{(0)},k_0)+o(1)\right),\quad
x\in\Omega^{ex}\backslash S^{ex}(\varepsilon^{1/2})
\endalign
$$ where  $X$ is the function defined in Lemma 2.2.
\endproclaim

I arises from (1.7) and (1.8) that in the case considered, the
two peak regimes of the solution of the boundary value problem
(1.1), (1.2)  "having the same rights" reads as follows:
$$
k=k(\varepsilon)=k_0+\varepsilon^{1/2}\tau_{1,0}^{(n)}+
\varepsilon\ln\varepsilon\,\,\tau_{2,1}^{(n)}+\varepsilon^2(t+o(1))
\tag1.10
$$ where $t$ is any real number. Substituting (1.10)
and the asymptotics from item c) of the theorem into (1.9), we
obtain that, for such $k$ the solution of the scattering problem
obeys asymptotics
$$ \allowdisplaybreaks \align u_\varepsilon(x;k) &
\sim{1\over\varepsilon^{1/2}}c^{(n)}_F(t) \psi(x),\quad
x\in\Omega^{in}\backslash S(\varepsilon^{1/2}),\\
u_\varepsilon(x;k) &
\sim{1\over\varepsilon^{1/2}}c^{(n)}_F(t)\psi(0),\qquad x\in
S(2\varepsilon^{1/2}),\\ u_\varepsilon(x;k) &
\sim{1\over\varepsilon}(-1)^nc^{(n)}_F(t) \left({2\over
h|\omega|}\right)^{1/2}\sin(k_0x_2),\qquad
 x\in\varkappa_\varepsilon\backslash(S^{ex}(\varepsilon^{1/2})
\cup S(\varepsilon^{1/2})),\\ u_\varepsilon(x;k) & \sim
c^{(n)}_F(t)(-1)^{m+n+1}{k_0\over\pi}\left({2|\omega|\over
h}\right)^{1/2} \left(\ln\varepsilon+
{\pi\over|\omega|}X\left({\left(x-x^{(0)}\right)^*\over\varepsilon}
\right)\right),\\ &\hskip7cm
 x\in S^{ex}(2\varepsilon^{1/2}),\\
u_\varepsilon(x;k) & \sim c^{(n)}_F(t)
(-1)^{m+n}k_0\left({2|\omega|\over h}\right)^{1/2}
G^{ex}(x,x^{(0)},k)+u^{ex}(x;k),\\ & \hskip7cm
x\in\Omega^{ex}\backslash S^{ex}(\varepsilon^{1/2}),\\ &
c^{(n)}_F(t)={(-1)^{m+n+1}\over2(t-\tau^{(n)}_{2,0})}
\left({|\omega|\over2h}\right)^{1/2}u^{ex}(x^{(0)};k_0).
\endalign
$$ These formulas show that an
interior resonance takes place  in  $\Omega^{in}$ in both cases
but it differs from (1.3) at order.

\head{\S~2. Construction of asymptotics}
\endhead

Denote $$ \allowdisplaybreaks \align R^{in(ex),i,j}_t(D_y)&=
\sum^t_{q=0}a^{in(ex)}_{i,j,q}{\partial^q\over\partial y_1^q},\\
\psi^{in(ex)}_\varepsilon(x,k)&=
\sum^\infty_{i=0}\sum^i_{j=0}\varepsilon^{i/2}
\ln^j\varepsilon(k^2_0-k^2)
R^{in(ex),i,j}_{[(i-j)/2]}(D_y)G^{in(ex)}(x,x^{in(ex)}_0,k),
\endalign
$$ where $x_0^{in}$ is the origin, $x^{ex}_0=x^{(0)}$,
$a^{in(ex)}_{i,j,q}$ are constants, $[N]$ is the integral part
of a number $N$. The coefficients of
$\psi^{in(ex)}_\varepsilon(x,k)$ are analytical with respect to
$k$ in some complex neighborhood of the point  $k_0$, satisfy
the equation $(\Delta+k^2)U=0$ in $\Omega^{in(ex)}$ and the
boundary condition $\partial U/\partial\nu=0$ on
$\partial\Omega^{in(ex)} \backslash \{x^{in(ex)}_0\}$. For
positive $k$, the coefficients of series $\psi
^{ex}_\varepsilon(x,k)$ also satisfy the radiation condition
(1.2). Therefore, outside the connecting channel and small
neighborhoods of its edges, following to [4], we seek the
complete asymptotics of the "eigenfunctions"
$\Psi_\varepsilon^{(n)}$ as
$$ \psi_\varepsilon^{(n)}(x) =
\psi^{in(ex)}_\varepsilon(x,\tau_\varepsilon^{(n)}),\qquad
x\in\Omega^{in(ex)}\backslash
S^{in(ex)}(\varepsilon^{1/2}),\tag2.1 $$ where $S^{in}(t)=S(t)$,
and  $R^{in(ex),i,j}_t$ in the definition of
$\psi^{in(ex)}_\varepsilon$ depend on $n$.
\par
\remark{Remark 2.1} The function $\Psi_\varepsilon^{(n)}$ in
(1.9) is not normalized, by $\psi_\varepsilon^{(n)}$ in (2.1) we
mean the function equalling $\Psi_\varepsilon^{(n)}$ up to
scalar factor $\alpha_n(\varepsilon)$. In \S~3, it will be shown
that $\alpha_n(\varepsilon)=1+o(1)$ as
$\varepsilon\rightarrow0$.
\endremark

\remark{Remark 2.2} The constructions of asymptotics for and for
associated generalized eigen\-func\-tions are identical. Because
of this and in order not to overload the notations in the text
by the index of correspondence to a concrete pole ($"n"$) we
will omit this index where it will be possible (including the
notations introduced above).
\endremark

In a small neighborhood of the connecting channel, asymptotics
of $\psi_\varepsilon^{(n)}$ are constructed in the form:
$$
\allowdisplaybreaks
\align
\psi_\varepsilon(x)
&
=\sum^\infty _{i=-1}
\sum^i_{j=0}\varepsilon^{i/2}\ln^j\varepsilon w_{i,j}(x_2),\qquad
x\in\varkappa_\varepsilon\backslash(S^{ex}(\varepsilon^{1/2})\cup
S^{in}(\varepsilon^{1/2})),\tag2.2\\
\psi_\varepsilon(x)
&
=\sum^\infty_{i=1-\beta_{in(ex)}}\sum^i_{j=0}
\varepsilon^{i/2}\ln^j\varepsilon v^{in(ex)}_{i,j}\left({x^{in(ex)}\over
\varepsilon}\right),\qquad
x\in S^{in(ex)}(2\varepsilon^{1/2}),\tag2.3
\endalign
$$ where $x^{in}=x$, $x^{ex}=(x-x^{(0)})^*$, $\beta_{in}=1$,
$\beta_{ex}=0$. Let us clarify the form of the leading terms in
(2.1) and (2.2). The formal limit of
$\psi^{in}_\varepsilon(x,k)$ as $\varepsilon\to0$ and $k\to k_0$
implies that $$ \psi^{in}_\varepsilon(x,k) \to R^{in,0,0}_0
\psi(0)\psi(x).\tag2.4
$$ On the other hand, in [4] it was shown that for
$k_0\in\Sigma^{in}_1\backslash\Sigma^{ch}$, the generalized
eigenfunction converges to the eigenfunction $\psi$ (continued
by zero outside $\Omega^{in}$), and for
$k_0\in\Sigma^{ch}\backslash\Sigma^{in}$, "in principle" it is
represented by
$$
{1\over\varepsilon^{1/2}}\left({2\over|\omega|h}\right)^{1/2}\sin(k_0t)\tag2.5
$$ in $\varkappa_\varepsilon$ and by zero outside
$\varkappa_\varepsilon$. Therefore, for
$k_0\in\Sigma^{in}_1\cap\Sigma^{ch}$, it is naturally to expect
that the leading term of the generalized eigenfunction is a
"linear" combination of  $\psi$ and (2.5). By the latter we
arrive at the leading terms indicated in (2.1), (2.2), and, in
particular an equality
 $$
w_{-1,0}(t)=b_{-1,0}\sin(k_0t).\tag2.6
$$ Moreover, $\psi$ being normalized, and the norm of
(2.5) in $L_2(\varkappa_\varepsilon)$ equaling  one, too, by
(2.4), (2.6) and by an assumption  (that will be justified
below) the conserving of the normalization for the leading term
of the generalized eigenfunction's asymptotics, for the case
considered in the paper, we obtain the equality
$$
\left(R_0^{in,0,0}\psi(0)\right)^2+{b^2_{-1,0}|\omega|h\over2}=1.\tag2.7
$$

The boundary value problems for the coefficients of the series
(2.3) are derived by a standard substitution ([4,7]) the series
 (1.7), (2.3) into (1.1) for $F=0$, and by passing to the
 "interior" variable $\xi=x^{in(ex)}\varepsilon^{-1}$: $$ \Delta
v_{i,j}=-k^2_0v_{i-2,j}-\sum^{i-4}_{q=1}
\sum^{q-1}_{t=0}\lambda_{q,t}v_{i-q-4,j-t},\quad\xi\in\gamma_\omega,\qquad
{\partial v_{i,j}\over\partial \nu} = 0,\quad \xi\in\partial
\gamma _\omega,\tag2.8 $$ where $\lambda _{q,t}$ denotes the
coefficients of the series
$\lambda_\varepsilon=\tau^2_\varepsilon-k^2_0$ in front of
$\varepsilon^{q/2}\ln^t\varepsilon$, $$
\gamma_\omega=\left(\omega\times(-\infty,0]\right)\cup\{\xi:\,\xi_2>0\},
$$ and the upper indexes for  $v^{in(ex)}_{q,t}$ are omitted.

Since the coefficients of the series (2.2) depends on $x_2$ only
(but considered as  functions defined on
$\varkappa_\varepsilon$),then substituting the series (1.7) and
(2.2) in (1.1), for $F=0$, we obtain the ordinary differential
equations for $w_{j,i}$:
$$
w^{\prime\prime}_{i,j}(x_2)+k^2_0w_{i,j}(x_2)+ \sum^{i+1}_{q=1}
\sum^{q-1}_{t=0}\lambda _{q,t}w_{i-q,j-t} (x_2) = 0,\qquad
-h<x_2<0, $$ whose solutions are functions
$$ \aligned w_{i,j}(x_2)=& -{1\over k_0}
\sum^{i+1}_{q=1}\sum^{q-1}_{t=0}\lambda_{q,t}
\int^{x_2}_{-h}\sin\left(k_0\left(x_2-t\right)\right)
w_{i-q,j-t}(t)\,dt\\ & +b_{i,j}\cos(k_0x_2),\qquad
j\ge0,\endaligned\tag2.9 $$ where $b_{i,j}$  are arbitrary
constants.  Hereafter, the coefficients of the series (2.2) are
chosen  in accordance with  (2.9). One can see, that in this
case
$$ w_{0,0}(t)= \tau_{1,0}k_0b_{-1,0}
\left(t\cos(k_0t)+h\cos(k_0t)-{1\over k_0}\sin(k_0t)\right)
+b_{0,0}\cos(k_0t). \tag2.10 $$

Let $\rho=\vert\xi\vert$, $P_j(\xi)$ be homogeneous polynomials
of order $j$, and $T_j(\xi)$ be homogeneous functions of order
$j$ represented as $P_{j+2q}(\xi)\rho^{-2q}$ for some integer
$q\ge1$ and satisfying the boundary condition $\partial
T_j(\xi)/\partial\xi_2=0$ as $\xi_2=0$, $\xi\not=0$. Denote by
$\tilde\Cal A_j$ the set of series of the form
$$ T(\xi)=\sum^j_{q=-\infty }T_q(\xi)+\ln\rho
\sum^j_{n=0}P_q(\xi).
$$ Next, we indicate by  $w_\varepsilon(x)$ the series (2.2).
Let us define the "re-extension" operator  $K^{in(ex)}_q$ for
the summation  $U(x,\varepsilon)$ of the form $w_\varepsilon(x)$
and $\psi^{in(ex)}_\varepsilon(x,\tau_\varepsilon)$, where
$\tau_\varepsilon$ is an arbitrary function with the asymptotics
(1.7), in the following standard way ([7]). We expand the
coefficients of $U(x,\varepsilon)$  in powers of  $\vert
x^{in(ex)}\vert\rightarrow0$   and pass to the variables
$\xi=x^{in(ex)}\varepsilon^{-1}$ (if
$U(x,\varepsilon)=\psi^{in(ex)}_\varepsilon (x,\tau
_\varepsilon)$, the function $\tau _\varepsilon$ is replaced by
its asymptotics series). In the double series obtained, we take
the sum of terms $\varepsilon^j\ln^i\varepsilon\Phi(\xi)$ for
$j\le q$. Exactly this sum is denoted by
$K^{in(ex)}_q(U(x,\varepsilon))$.

We indicate  $$\allowdisplaybreaks\align & g^{in}(k)=\lim_{x\to
0} \left(G^{in}(x,0,k)+{1\over\pi}\ln r+ {\psi^2(0)\over
k^2-k^2_0}\right),
\\ & g^{ex}(k)=\lim_{x\to x^{(0)}}
\left(G^{ex}(x,x^{(0)},k)+{1\over\pi}\ln|x-x^{(0)}|\right),
\endalign$$
and by $(\rho,\theta)$ we denote the polar coordinates. The
definitions of  $\psi^{in(ex)}_\varepsilon$, $w_\varepsilon$,
$K^{in(ex)}_{q}$, the equalities (2.6) and (2.10) and the
asymptotics of the Green functions and of their derivatives
(see, for instance,  [14]) lead as to

\proclaim{Lemma 2.1}  Let $\tau_\varepsilon$ be an arbitrary
function with asymptotics  (1.7). Then for any integer $N\ge0$
the equalities
$$ \allowdisplaybreaks \aligned
K^{in(ex)}_{N/2}( \psi^{in(ex)}_\varepsilon(x,\tau_\varepsilon))
& =\sum^N_{i=1-\beta_{in(ex)}}\sum^i_{j=0}
\varepsilon^{i/2}\ln^j\varepsilon V^{in(ex)}_{i,j}(\xi),\\
K^{in(ex)}_{N/2}(w_\varepsilon(x)) & =\sum^N_{i=0}\sum^i_{j=0}
\varepsilon^{i/2}\ln^j\varepsilon W^{in(ex)}_{i,j}(\xi_2),
\endaligned
$$
hold, where $V^{in(ex)}_{i,j}\in\tilde\Cal A_{[(i-j)/2]}$, and
$W^{in(ex)}_{i,j}$ are polynomials $[(i+1-j)/2]$-th order.

The representation
$$
\allowdisplaybreaks
\align
V^{in}_{0,0}(\xi)=
&
R^{in,0,0}_0\psi^2(0),\\
V^{in(ex)}_{1,0}(\xi)=
&
\beta_{in(ex)}R^{in(ex),1,0}_0\psi^2(0)\\
&
+
{2\over\pi}k_0\tau_{1,0}\left(R^{in(ex),0,0}_0
\left(\ln\rho-\pi g^{in(ex)}(k_0)\right)-\Pi_{0,0}\right),\\
V^{in(ex)}_{k,k}(\xi)
=&
{2k_0\over\pi}\sum_{t=0}^{k-1}R^{in(ex),t,t}_0\tau_{k-t,k-t-1},
\qquad k\ge1,\\
V^{in(ex)}_{k,j}(\xi)=
&
\tilde V^{in(ex)}_{k,j}(\xi)+
\beta_{in(ex)}a^{in(ex)}_{k,j,0}\psi^2(0)
\\
& \hskip-1cm+
{2\over\pi}k_0\Bigg(\tau_{1,0}\left(a^{in(ex)}_{k-1,j,0}
\left(\ln\rho-\pi g^{in(ex)}(k_0)\right)-\Pi_{k-1,j}\right)\\ &
\hskip-1cm+ \tau_{k,j}\left(R^{in(ex),0,0}_0 \left(\ln\rho-\pi
g^{in(ex)}(k_0)\right)-\Pi_{0,0}\right)\Bigg),\quad k>1,\,\,k>j,\\
\Pi_{i,j} & =\sum_{t=1}^\infty
a^{in(ex)}_{2t+i,j,t}(t-1)!{\cos(t\theta)\over \rho^{t}},\\ \tilde
V^{in}_{2,0}(\xi) &
=
R^{in,0,0}_0\psi(0)\psi_{x_1}(0)\xi_1,\qquad
\tilde V^{ex}_{2,0}(\xi)=\tilde V^{in(ex)}_{2,1}(\xi)=0,
\endalign
$$
are valid, where $\tilde V^{in(ex)}_{i,j}$ are independent on
$\tau_{q,s}$, $\Pi_{q-1,s}$ and $a^{in(ex)}_{q-1,s,0}$ for $q\ge
i$ and $s\ge j$,
$$
\allowdisplaybreaks
\align
W^{in}_{i,i}(\xi_2)
&
=\tau_{i+1,i}b_{-1,0}h+b_{i,i},\qquad
W^{ex}_{i,i}(\xi_2)
=(-1)^mb_{i,i},\\
W^{in}_{k,j}(\xi_2)
&
=\tilde W^{in}_{k,j}(\xi_2)+
b_{-1,0}h\left({1\over2}\tau_{k,j}\tau_{1,0}
+\tau_{k+1,j}\right)
+b_{k,j},\\
W^{ex}_{k,j}(\xi_2)
&
=\tilde W^{ex}_{k,j}(\xi_2)+(-1)^mb_{k,j},\qquad k>j,\\
\tilde W^{in}_{1,0}(\xi_2)
&
=b_{-1,0}k_0\xi_2,\qquad
\tilde W^{ex}_{1,0}(\xi_2)=(-1)^{m+1}b_{-1,0}k_0\xi_2,\\
\tilde W^{in}_{k+1,k}(\xi_2)
&
=\alpha^{in}_k,\qquad \tilde W^{ex}_{k+1,k}(\xi_2)=\alpha^{ex}_k,
\qquad k\ge1,\qquad
\tilde W^{in(ex)}_{2,0}(\xi_2)=0,
\endalign
$$
where $\tilde W^{in(ex)}_{i,j}$ are independent on $\tau_{q,s}$
and $b_{q,s}$, for $q\ge i$ and $s\ge j$, and
$\alpha^{in(ex)}_k$ are some constants.

The series  $V^{in(ex)}_{i,j}$  (and, hence, the series $\tilde
V^{in(ex)}_{i,j}\in\Cal A_{[(i-j)/2]}$) are formal asymptotic
solutions to the boundary problem (2.8) for
$\rho\rightarrow\infty$, $\xi_2\ge0$, where the functions
$v_{q,t}$ are replaced by $V^{in(ex)}_{q,t}$.

If $W^{ex}_{0,0}\equiv0$, then $W^{in(ex)}_{i,j}$ (therefore,
and $\tilde W^{in(ex)}_{i,j}$) are formal asymptotic solutions
to the boundary problem (2.8)  for $\rho \rightarrow \infty $,
$\xi_2<0$, where the functions $v_{q,t}$ are replaced by
$W^{in(ex)}_{q,t}$.
\endproclaim

In order to match the series (2.1)--(2.3), it is sufficient to
show the existence of the solutions to the boundary value
problems (2.8), whose asymptotics at infinity are
$V^{in(ex)}_{i,j}$ for $\xi_2\ge0$ and $W^{in(ex)}_{i,j}$ for
$\xi_2<0$. Let $0=\mu_0<\mu_1\le\mu_2\le...$ be the
eigenfrequencies of the Neumann problem for the operator
$-d^2/dt^2$ in the interval  $\omega$, $\beta_q(t)$ be the
associated eigenfunctions normalized in $L_2(\omega)$. We ill
employ the symbol $\tilde\Cal B_{q,n}$ for the set of series
$$
H(\xi)=R^{(0)}_q(\xi_2)+\sum^\infty_{j=1}R^{(j)}_n
(\xi_2)\beta_j(\xi_1)\exp\{\mu_j\xi_2\},
$$
where $R^{(m)}_i(t)$ are polynomials of order $i$. For negative
$n$, $\tilde\Cal B_{q,n}$ denotes the set of po\-ly\-no\-mials
of order $q$. We set $\tilde\Cal B =\cup_{n,q}\tilde\Cal
B_{q,n}$. We denote by $\Cal A_m$ the set of functions from
$C^\infty(\gamma_\omega)\cap W^1_{2,loc}(\gamma_\omega)$
satisfying the homogeneous Neumann boundary condition on
$\partial\gamma_\omega$ and having differentiable asymptotics
from $\tilde\Cal A_m$ and $\tilde\Cal B$ at infinity for
$\xi_2\ge0$ and $\xi_2<0$, respectively. Matching of the series
will be proved by using Lemma 2.1 and the following statement
proved in [4].

\proclaim{Lemma 2.2}  Let  $f\in\Cal A_N$ and the series
$V\in\tilde\Cal A_{N+2}$  be the formal asymptotic solution of
the equation $\Delta V=f$ for $\rho\rightarrow\infty $,
$\xi_2\ge0$, and the polynomials $W(\xi_2)$ satisfy the equation
$\Delta W=f+o(1)$ (or the equation $W^{\prime\prime}=f+o(1)$,
what is the same) for $\rho\rightarrow\infty$ and $\xi_2<0$.

Then, there exists a function $v\in {\Cal A}_{N+2}$ that is the
solution of the boundary value problem
$$ \Delta v=f,\quad\xi\in\gamma_\omega,\qquad {\partial
v\over\partial \nu} = 0,\quad\xi\in\partial\gamma_\omega\tag2.11
$$
and has the following differentiable asymptotics as
$\rho\rightarrow\infty$

$$ \allowdisplaybreaks \aligned v(\xi)& = \tenrm
V(\xi)+ñ_0\ln\rho+\sum^\infty_{j=1}ñ_j {\cos(j\theta)\over
\rho^{j}} ,\qquad\xi_2\ge0,\\ v(\xi)&
=W(\xi_2)+q_0+O(\xi^M_2\exp\{\mu_1\xi_2\}),\qquad\xi_2<0,
\endaligned
$$
where $M\ge0$, $q_0$ and  $c_i$ are some numbers.

There exist functions $X\in\Cal A_0$ and $Y\in\Cal A_1$ that is
harmonic in  $\gamma_\omega$ and have the following
differentiable asymptotics at infinity:
$$ \aligned X(\xi)&
=\xi_2+q_\omega+O\left(\exp\{\mu_1\xi_2\}\right), \qquad
\xi_2\le0,\\ X(\xi)& =c_\omega\ln \rho +
\sum^\infty_{j=1}c^+_j{\cos(j\theta)\over \rho^j},\qquad
\xi_2\ge0,\\ Y(\xi)& =\xi_1+c^\omega\ln\rho
+\sum^\infty_{j=1}b_j{\cos(j\theta)\over \rho^j},\qquad
\xi_2\ge0,\\ Y(\xi)&
=q^\omega+O(\exp\{\mu_1\xi_2\}),\qquad\xi_2<0.
\endaligned
$$
\endproclaim

For $\omega_-=-\omega_+$, taking in account evenness, one can
see that the constants $q^\omega$ and $c^\omega$ equal zero.
Thus, it is clear that, in the general (nonsymmetric) case:
$$ c^\omega=0,\qquad q^\omega={\omega_++\omega_-\over2}.
$$
It is easy to establish the equalities
$c_\omega=\pi^{-1}|\omega|$, $q_\omega=\pi^{-1}|\omega|
\left(\ln\left(2|\omega|\pi^{-1}\right)-1\right)$ by using
conformal mapping of a strip onto  $\gamma_\omega$ (see, for
instance, [4,15]).

We denote by $v^{in(ex)}_\varepsilon(x^{in(ex)}/\varepsilon)$
the series  (2.3), and by  $v^{in(ex)}_{\varepsilon,N}(\xi)$
their partial sum and come to prove the key statement of the
present work.

\proclaim{Theorem 2.1} There exist a function $\tau_\varepsilon$
with asymptotics (1.7) and series  (2.1)--(2.3) such that the
coefficients $v^{in(ex)}_{i,j}\in{\Cal A}_{[(i-j)/2]}$ are the
solutions of the recurrent boundary value problems (2.8), the
coefficients $w_{i,j}$ are defined by (2.6), (2.9), and for any
integer  $N\ge0$ the following differentiable asymptotic
equalities  hold:
$$ \allowdisplaybreaks \align &
K^{in(ex)}_{N/2}(w_\varepsilon(x))
=v^{in(ex)}_{\varepsilon,N/2}(\xi)+O(\varepsilon^{N/2}\xi^{M_N}_2
\exp\{\mu_1\xi_2\}),\qquad \xi_2<0,\tag2.12\\ &
K^{in(ex)}_{N/2}(\psi^{in(ex)}_\varepsilon(x,\tau_\varepsilon))
= v^{in(ex)}_{\varepsilon,N/2}(\xi),\qquad \xi_2\ge0,\qquad
\rho\to\infty, \tag2.13
\endalign
$$ and for the coefficients of these series the following
representations are true:
$$ \allowdisplaybreaks \align &
b_{0,0}=0,\qquad v^{in}_{0,0}\equiv R^{in,0,0}_0\psi^2(0),\qquad
\qquad R^{ex,0,0}_0=(-1)^{m+1}R^{in,0,0}_0,\\ &
b_{-1,0}=\left({1\over h|\omega|}\right)^{1/2},\qquad
R^{in,0,0}_0=\pm{1\over\psi(0)}
\left({1\over2}\right)^{1/2},\qquad
\tau_{1,0}=\pm\psi(0)\left({|\omega|\over2h}\right)^{1/2},\\ &
v^{ex}_{1,1}\equiv {2k_0\tau_{1,0}\over\pi}R^{ex,0,0}_0,\qquad
\tau_{2,1}={4k_0\over b_{-1,0}h\pi}R^{in,0,0}_0,\tag2.14 \\ &
\tau_{2,0}={\tau_{1,0}\over
\tau_{1,0}b_{-1,0}h+R^{in,0,0}_0\psi^2(0)}\times\\ &\hskip3em
\left(b_{-1,0}\left(2k_0q_\omega-{1\over2}h\tau^2_{1,0}\right)
-2k_0\tau_{1,0}R^{in,0,0}_0\left(g^{in}(k_0)+g^{ex}(k_0)\right)
\right)
\endalign
$$
\endproclaim

\demo{Proof} Setting $b_{0,0}=0$, due to Lemma 2.1, we achive
the equalities (2.12), (2.13) for the index  $"ex"$ as
 $N=0$. Similarly, taking $v^{ex}_{0,0}$ as in  (2.14) and
 putting the additional condition
$$
R^{in,0,0}_0\psi^2(0)=\tau_{1,0}hb_{-1,0},\tag2.15
$$
where $R^{in,0,0}_0$, $\tau_{1,0}$ and $b_{-1,0}$ are some
constants unknown yet but satisfying (2.7), due to Lemma 2.1
(and also due to value of $b_{0,0}$ defined above) we get the
equalities (2.12), (2.13) for the index $"in"$ as $N=0$. This
was the "zero step" of the matching procedure.

At the next step, by virtue of definition $\tilde
W^{in,ex}_{1,0}$ è $V^{in,ex}_{1,0}$, we put
$$
\allowdisplaybreaks
\align
&
v^{in}_{1,0}=b_{-1,0}k_0X+A^{in}_{1,0}, \\
&
v^{ex}_{1,0}=(-1)^{m+1}b_{-1,0}k_0X+A^{ex}_{1,0},
\endalign
$$
where $A^{in(ex)}_{1,0}$ are some constants unknown yet. It is
easy to see that these functions are the solutions of the
boundary value problem (2.8). Setting the (power) asymptotics of
the functions $v^{in(ex)}_{1,0}$ as $\rho\to\infty$ and
$\xi_2<0$ equal to $W^{in(ex)}_{1.0}$, due to Lemma 2.1 we
obtain the equations
$$
\allowdisplaybreaks
\align
& (-1)^{m+1}b_{-1,0}k_0q_\omega+A^{ex}_{1,0}=(-1)^mb_{1,0},\\ &
b_{-1,0}k_0q_\omega+A^{in}_{1,0}=b_{-1,0}h\left({1\over2}\tau^2_{1,0}
+
\tau_{2,0}\right)+b_{1,0},\tag2.16
\endalign
$$
Similarly, setting the asymptotics of the functions
$v^{in(ex)}_{1,0}$ as $\rho\to\infty$ and $\xi_2>0$ equal to the
series $V^{in(ex)}_{1.0}$ up to the terms  $O(1)$ inclusive, we
obtain the equations
$$
\allowdisplaybreaks
\align
&
b_{-1,0}k_0c_\omega={2\over\pi}k_0\tau_{1,0}R^{in,0,0}_0,\tag2.17\\
&
(-1)^{m+1}b_{-1,0}k_0c_\omega={2\over\pi}k_0\tau_{1,0}R^{ex,0,0}_0,
\tag2.18\\
&
R^{in,1,0}_0\psi^2(0)-2k_0\tau_{1,0}R^{in,0,0}_0g^{in}(k_0)=
A^{in}_{1,0},\tag2.19\\
&
-2k_0\tau_{1,0}R^{ex,0,0}_0g^{ex}(k_0)=A^{ex}_{1,0},\tag2.20
\endalign
$$
where $R^{in,1,0}_0$ is one more unknown constant. From (2.17)
and (2.18) we get the value (2.14) for $R^{ex,0,0}_0$, while
(2.20) determines the constant $A^{ex}_{1,0}$. Solving the
system of the equations (2.7), (2.15), (2.17) we get the
formulae (2.14) for $R^{in,0,0}_0$, $b_{-1,0}$ and $\tau_{1,0}$.
We stress the calculating  the latter quantities finally
determines the constants $R^{ex,0,0}_0$, $A^{ex}_{1,0}$,
$$
b_{1,0}=k_0\left(2\tau_{1,0}R^{in,0,0}_0g^{ex}(k_0)-b_{-1,0}q_\omega\right)
\tag2.21
$$
and the function $v^{ex}_{1,0}$. Also, the function
$v^{in}_{1,0}$ is determined up to the additive term
$A^{in}_{1,0}$, which satisfies equation (2.17). Moreover,
setting the asymptotics of the functions $v^{in(ex)}_{1,0}$ as
$\rho\to\infty$ and $\xi_2>0$ equal to the series
$V^{in(ex)}_{1.0}$ for the other degree, we determine
$\Pi^{in(ex)}_{0,0}$ (i.e., the coefficients of the higher
derivatives for the polynomials $R^{in(ex),2i,0}_i$). And,
finally, putting
$$
v^{in}_{1,1}\equiv {2k_0\over\pi}R^{in,0,0}_0\tau_{1,0},
$$
and defining $v^{ex}_{1,1}$ in accordance with (2.14), and
setting the "asymptotics" of these functions  equal to the
"series" $V^{in(ex)}_{1,1}$ and to the "polynomials"
$W^{in(ex)}_{1,1}$ as $\xi_2>0$ and $\xi_2<0$, respectively, we
obtain the equalities (2.12) and (2.13) for $N=1$, and get the
value (2.14) for $\tau_{2,1}$ (simultaneously determining
$b_{1,1}$). This was the first step of the matching procedure,
which in addition to the equalities  (2.12), (2.13) as  $N=1$
gives two equations (2.16) and (2.19) to three constants
$R^{in,1,0}_0$, $A^{in}_{1,0}$ and $\tau_{2,0}$ unknown yet.

In the second step, by $\tilde V^{in(ex)}_{2,0}$ and $\tilde
W^{in(ex)}_{2,0}$, we determine the solutions $v^{in(ex)}_{2,0}$
of the boundary value problem (2.8) as
$$
v^{in}_{2,0}=R^{in,0,0}_0\psi(0)\psi_{x_1}(0)Y+A^{in}_{2,0},\qquad
v^{ex}_{2,0}\equiv A^{ex}_{2,0}, $$ where $A^{in(ex)}_{2,0}$ are
some constants. Setting the asymptotics of the functions
$v^{in(ex)}_{2,0}$ at infinity as $\xi_2<0$ equal to
$W^{in(ex)}_{2,0}$, we obtain the equalities:
$$ \allowdisplaybreaks \align &
A^{in}_{2,0}+R^{in,0,0}_0\psi(0)\psi_{x_1}(0)q^\omega=b_{-1,0}h
\left({1\over2}\tau_{2,0}\tau_{1,0}
+\tau_{3,0}\right)+b_{2,0},\tag2.22\\ &
A^{ex}_{2,0}=(-1)^mb_{2,0}.\tag2.23
\endalign
$$
Similarly, setting the asymptotics of the functions
$v^{in(ex)}_{2,0}$ at infinity as  $\xi_2>0$ equal to
$V^{in(ex)}_{2,0}$ up to terms $O(1)$, we get the following
equalities:
$$
\allowdisplaybreaks
\align
&
\tau_{1,0}R^{in,1,0}_0+
\tau_{2,0}R^{in,0,0}_0=0,\tag2.24\\
&
\tau_{1,0}R^{ex,1,0}_0+
\tau_{2,0}R^{ex,0,0}_0=0,\tag2.25\\
&
-
2k_0g^{ex}(k_0)\left(\tau_{1,0}R^{ex,1,0}_0+
\tau_{2,0}R^{ex,0,0}_0\right)=A^{ex}_{2,0},\tag2.26\\
&
a^{in}_{2,0,0}\psi^2(0)-
2k_0g^{in}(k_0)\left(\tau_{1,0}R^{in,1,0}_0+
\tau_{2,0}R^{in,0,0}_0\right)=A^{in}_{2,0}.\tag2.27
\endalign
$$
From (2.25) and (2.26) we obtain   $A^{ex}_{2,0}$ that, due to
(2.23), determines $b_{2,0}$. Furthermore, solving the system of
the equations (2.16), (2.19), (2.24) (bearing in mind the
equality (2.21), too), we determine $\tau_{2,0}$ in accordance
with (2.14), and also get the constants  $R^{in,1,0}_0$ and
$A^{in}_{1,0}$. The coefficients  $R^{in,1,0}_0$ and
$\tau_{2,0}$ having been determined, first, from (2.26) we
obtain $R^{ex,1,0}_0$, and, second, setting the asymptotics of
the functions $v^{in(ex)}_{2,0}$ at infinity as $\xi_2>0$ equal
to $V^{in(ex)}_{2,0}$ (for the other terms) we define all
coefficients of the series $\Pi_{1,0}^{in(ex)}$ (i.e., the
leading coefficients of the differential polynomials
$R^{in(ex),1+i,0}_i$ as $i\ge1$). The equations (2.22), (2.27)
are analogue of the equations (2.16) and (2.19) for determining
(in the next step) the constants $A^{in}_{2,0}$, $\tau_{3,0}$
and $a^{in}_{2,0,0}$.

The subsequent proof is carried out by induction. Before
beginning the $(N,k)$-th step, where $k$ is a degree of
$\ln\varepsilon$, all $\Pi^{in(ex)}_{q-2,s}$, $\tilde
W^{in(ex)}_{q,s}$, $\tilde V^{in(ex)}_{q,s}$
$a^{in(ex)}_{q-1,s,0}$, $\tau_{q-1,s}$, $v^{in(ex)}_{q-2,s}$,
$v^{ex}_{q-1,s}$ as $q\le N$ and $s\le k$ have been determined,
and the functions $v^{in}_{N-1,s}$ are determined up to additive
terms $A^{in}_{N-1,s}$, meeting the following equations
(analogue of (2.22) and (2.27)):
$$
\allowdisplaybreaks
\aligned
&
A^{in}_{N-1,s}+B^{in}_{N-1,s}=b_{-1,0}h\tau_{N,s},\\
&
a^{in}_{N-1,s,0}\psi^2(0)+C^{in}_{N-1,s}=A^{in}_{N-1,s},\qquad s\le N-1,
\endaligned\tag2.28
$$
where $B^{in}_{N-1,s}$ and $C^{in}_{N-1,s}$ are some completely
defined numbers. In the $(N,s)$-th step, where $s<N$, by $\tilde
W^{in(ex)}_{N,s}$ and $\tilde V^{in(ex)}_{N,s}$, we determine
$v^{in(ex)}_{N,s}$ as
$$
v^{in(ex)}_{N,s}=\tilde v^{in(ex)}_{N,s}+A^{in(ex)}_{N,s},
$$
where $A^{in(ex)}_{N,s}$ are some undetermined  constants and
$\tilde v^{in(ex)}_{N,s}$ are the solutions of the boun\-da\-ry
value problems (2.8) with the asymptotics:
$$
\allowdisplaybreaks
\align
&
\tilde v^{in(ex)}_{N,s}=\tilde W^{in(ex)}_{N,s}+D^{in(ex)}_{N,s}+o(1),
\qquad \xi_2<0,\\
&
\tilde v^{in(ex)}_{N,s}=\tilde V^{in(ex)}_{N,s}+E^{in(ex)}_{N,s}\ln\rho+
F^{in(ex)}_{N,s}+O(\rho^{-1}),\qquad
\xi_2>0,
\endalign
$$
where $D^{in(ex)}_{N,s}$, $E^{in(ex)}_{N,s}$ and
$F^{in(ex)}_{N,s}$ are also completely  defined constants. The
existence the such functions follows from the statement of
Lemmas 2.1 and 2.2. Setting the asymptotics of the functions
$v^{in(ex)}_{N,s}$ as $\rho\to\infty$ and $\xi<0$ equal to the
polynomials $W^{in(ex)}_{N,s}$, we obtain two equations:
$$
\allowdisplaybreaks
\align
&
A^{in}_{N,s}+D^{in}_{N,s}=b_{-1,0}h
\left({1\over2}\tau_{N,s}\tau_{1,0}
+\tau_{N+1,s}\right)+b_{N,s},\tag2.29\\ &
A^{ex}_{N,s}+D^{ex}_{N,s}=(-1)^mb_{N,s}.\tag2.30
\endalign
$$
Similarly, setting the asymptotics of the functions
$v^{in(ex)}_{N,s}$ at infinity as  $\xi_2>0$ equal to the series
$V^{in(ex)}_{N,s}$ up to the terms $O(1)$, we get the
equalities:
$$
\allowdisplaybreaks
\align
&
{2k_0\over\pi}
\left(\tau_{1,0}a^{in}_{N-1,s,0}+
\tau_{N,s}R^{in,0,0}_0\right)=E^{in}_{N,s},\tag2.31\\
&
{2k_0\over\pi}
\left(\tau_{1,0}a^{ex}_{N-1,s,0}+
\tau_{N,s}R^{ex,0,0}_0\right)=E^{ex}_{N,s},\tag2.32\\
&
-
2k_0g^{ex}(k_0)\left(\tau_{1,0}a^{ex}_{N-1,s,0}
+
\tau_{N,s}R^{ex,0,0}_0\right)=A^{ex}_{N,s},\tag2.33\\
&
a^{in}_{N,s,0}\psi^2(0)-
2k_0g^{in}(k_0)\left(\tau_{1,0}a^{in}_{N-1,s,0}+
\tau_{N,s}R^{in,0,0}_0\right)=A^{in}_{N,s}.\tag2.34
\endalign
$$
From (2.32) and (2.33) we obtain  $A^{ex}_{N,s}$ what, due to
(2.30), defines $b_{N,s}$. Furthermore, solving the system of
the equations (2.28), (2.32), we get  $\tau_{N,s}$,
$a^{in}_{N-1,s,0}$ and $A^{in}_{N-1,s}$. The coefficients
$a^{in}_{N-1,s,0}$ and $\tau_{N,s}$ having been determined,
first, from (2.33) we obtain $a^{ex}_{N,s,0}$, and, second,
setting the asymptotics of the functions $v^{in(ex)}_{N,s}$ at
infinity as  $\xi_2>0$ equal to the series $V^{in(ex)}_{N,s}$
(for other terms), we get all coefficients of the series
$\Pi_{N-1,s}^{in(ex)}$. The equations (2.29), (2.34) are
analogue of (2.28) for the $(N+1,s)$-th step.

The same procedure repeats in the $(N,s+1)$-th step (if
$s+1<N$). In the  $(N,N)$-th step the situation is  simpler. In
this case, Lemma 2.1 implies that
$$
\aligned
&
v^{in}_{N,N}\equiv{2k_0\over\pi}
\sum_{t=0}^{N-1}R^{in,t,t}_0\tau_{k-t,k-t-1}=
\tau_{N+1,N}b_{-1,0}h+b_{N,N},\\
&
v^{ex}_{N,N}\equiv{2k_0\over\pi}
\sum_{t=0}^{N-1}R^{ex,t,t}_0\tau_{k-t,k-t-1}=(-1)^mb_{N,N}.
\endaligned\tag2.35
$$
Solving (2.35), we obtain $b_{N,N}$ è $\tau_{N+1,N}$. Theorem is
proved.
\enddemo

We stress that in Theorem 2.1 it is constructed two asymptotic
series corresponding to
$\tau_{1,0}=(-1)^{n}\psi(0)(|\omega|/2h)^{1/2}$, $n=1,\,2$ (or
$R^{in,0,0}_0=(-1)^{n}\psi_0^{-1}2^{-1/2}$, which is the same).
More\-over, from the formulae (2.14) it follows (in the formal
level, for now) the formulas (1.8) for $\tau_{1,0}$ and
$\tau_{2,1}$, and the equality $$ \tau_{2,0}^{(n)}= {1\over2}
{|\omega|\over h} \left(
{2k_0\over\pi}\left(\ln\left({2|\omega|\over\pi}\right)-1\right)
-{1\over4}\psi^2(0) -k_0\left(g^{in}(k_0)+g^{ex}(k_0)\right)
\right). $$ Employing this equality and taking in account that
$\hbox{\rm Im}\, g^{in}(k_0)=0$, $\hbox{\rm Im}\,
g^{ex}(k_0)=k_0\sigma$ (see, for instance, [14,16]), we obtain
the formula (1.8) for $\hbox{\rm Im}\, \tau_{2,0}$. Finally,
from (2.14) it follows that the leading terms of the series
(2.1)--(2.3) has the form indicated for the functions
$\Psi^{(n)}_\varepsilon$ in the statement c) of Theorem 1.1 in
the corresponding domains.

Denote $$ \allowdisplaybreaks \align
\psi_{\varepsilon,N}^{(n)}(x,k) &
=
\chi\left({|x^{in}|\over\varepsilon^{1/2}}\right)
\sum^N_{i=0}\sum^i_{j=0}\varepsilon^{i/2}
\ln^j\varepsilon(k^2_0-k^2)
R^{in,i,j}_{[(i-j)/2]}(D_y)G^{in}(x,x^{in}_0,k)\\ &
+\chi\left({|x^{ex}|\over\varepsilon^{1/2}}\right)
\sum^N_{i=0}\sum^i_{j=0}\varepsilon^{i/2}
\ln^j\varepsilon(k^2_0-k^2)
R^{ex,i,j}_{[(i-j)/2]}(D_y)G^{ex}(x,x^{ex}_0,k)\\ &
+\left(1-\chi\left({|x^{in}\over
|\varepsilon^{1/2}}\right)\right)v^{in}_{\varepsilon,N/2}
\left({x^{in}\over\varepsilon}\right)+
\left(1-\chi\left({|x^{ex}\over|\varepsilon^{1/2}}\right)
\right)v^{ex}_{\varepsilon,N/2}
\left({x^{ex}\over\varepsilon}\right)\\ & +
\chi\left({|x^{in}|\over\varepsilon^{1/2}}\right)
\chi\left({|x^{ex}|\over\varepsilon^{1/2}}\right)
w_{\varepsilon,N/2}(x),
\endalign
$$ where $w_{M,\varepsilon}$ is the partial sum of the series
(2.2), $\chi(t)$ is a smooth cut-off function equalling to one
for $t>2$ and vanishing  as $t<0$, and the index $n=1,2$ in the
left side corresponds as two series of the asymptotics and it is
omitted in the right side for sake of brevity. From Theorem 2.1
by standard way (see, for instance,  [17]) it follows

\proclaim{Corollary} Let the asymptotics of the function
$\tau_\varepsilon^{(n)}$ and the series (2.1)--(2.3) satisfy the
statements Theorem 2.1. Then

(a) $\psi_{\varepsilon,N}^{(n)}(x,k)\in
C^\infty(\Omega_\varepsilon)$ is a holomorphic function from
 $W^1_{2,loc}(\Omega_\varepsilon)$ which, for
 $\hbox{\rm Im}\,k\ge0$, satisfies (1.2);

(b) $\psi_{\varepsilon,N}^{(n)}(x_,k)$ is the solution of (1.1),
where $F(x,k)=F_{\varepsilon,N}^{(n)}(x,k)$ is a holomorphic
function from $L_2({\Bbb R}^2)$, $\hbox{\rm
supp}\,F_{\varepsilon,N}\subset\varkappa_\varepsilon\cup
S^{in}(2\varepsilon^{1/2})\cup S^{ex}(2\varepsilon^{1/2})$ and
$$
\parallel F_{\varepsilon,N}^{(n)}(\circ,\tau_\varepsilon^{(n)})
\parallel_{L_2({\Bbb
R}^2)} \le C_N\varepsilon^{N_1},
$$
where $N_1$ increases unboundedly with $N$.
\endproclaim

From the explicit form (2.14) of the leading terms of the
asymptotics also it follows that
$$
||\psi_{\varepsilon,N}^{(n)}(\bullet,\tau_\varepsilon^{(n)})||_
{L_2(\Omega_\varepsilon\cap S(T))}\to1,\quad
\int\limits_{\Omega_\varepsilon\cap
S(T)}\psi_{\varepsilon,N}^{(1)}(x, \tau_\varepsilon^{(1)})
\psi_{\varepsilon,N}^{(2)}(x,\tau_\varepsilon^{(2)}) \,dx\to
0\tag2.36
$$
as $\varepsilon\to0$ for each $T$ sufficiently large.

The formal construction of the asymptotics is finished.

\head{\S 3. Justification of the asymptotics}
\endhead

From [3] it follows

\proclaim{Lemma 3.1} There exist

(a) not more than two poles $\tau_\varepsilon^{(n)}$ converging
to $k_0\in\Sigma^{in}_1\cap\Sigma^{ch}$ as $\varepsilon\to0$;

(b) if $\tau_\varepsilon^{(1)}\not=\tau_\varepsilon^{(2)}$,
then, for each pole, there exist  only one generalized
eigenfunction.
\endproclaim

In its turn, following  [1,16] and using Lemma 3.1, it is easy
to prove the following statement.

\proclaim{Lemma 3.2} Let $k_0\in\Sigma^{in}_1\cap\Sigma^{ch}$,
$F\in L_2(\Bbb R^2)$ and $\hbox{\rm supp} F\subset S(R)$. Then

(a) for small $\varepsilon$ and $k$ close to $k_0$,
 the following uniform estimate
holds for the analytic continuation of the solution of the
boundary value problem (1.1), (1.2):
$$
||u_\varepsilon||_{L_2(S(T))}\le
{C(R,T)\over|(\tau_\varepsilon^{(1)}-k)(\tau_\varepsilon^{(2)}-k)|}
||F||_{L_2(\Bbb R^2)},
$$
which is independent of that poles $\tau_\varepsilon^{(n)}$
coincide or not;

(b) if $\tau_\varepsilon^{(1)}\not=\tau_\varepsilon^{(2)}$, then
the statement (b) of the theorem 2.1 is true, and, for small
$\varepsilon$ and $k\in K$ close to $k_0$, the following uniform
estimate holds
$$
||\tilde u_\varepsilon||_{L_2(S(T))}\le C(R,T)||F||_{L_2(\Bbb
R^2)};
$$

(c) if $\tau_\varepsilon^{(1)}\not=\tau_\varepsilon^{(2)}$ and
for $T>0$ sufficiently large

$$ {1\over\|\Psi^{(1)}_\varepsilon\|_{L_2(\Omega_\varepsilon\cap
S(T))} \|\Psi^{(2)}_\varepsilon\|_{L_2(\Omega_\varepsilon\cap
S(T))}} \int\limits_{\Omega_\varepsilon\cap
S(T)}\Psi^{(1)}_\varepsilon \Psi_\varepsilon^{(2)}\,dx\to
0,\quad \varepsilon\to0, $$ then
$||\Psi^{(n)}_\varepsilon||_{L_2(\Omega_\varepsilon\cap
S(R))}\to1$ as $\varepsilon\to0$.
\endproclaim

\demo{Proof of Theorem 1.1} Validity of the statement a) follows
from the statements (a) of Lemma 3.2, Corollary of Theorem 2.1
and the arbitrary choice of $N$. Since,
$\tau^{(1)}_\varepsilon\not=\tau^{(2)}_\varepsilon$ validity of
the statement b) follows from the statement (b) of Lemma 3.2.
Furthermore, from the statements (b) of Lemma 3.2, Corollary of
Theorem 2.1 and the arbitrary choice of $N$, it follows that in
representation (1.9): $$
\Psi_{\varepsilon}^{(n)}(x)=\alpha_n(\varepsilon)
\psi_{\varepsilon}^{(n)}(x),\tag3.1 $$ where
$\psi_{\varepsilon}^{(n)}$ has the asymptotics (2.1)--(2.3),
whose coefficients satisfy the statements of Theorem 2.1, and
$\alpha_n(\varepsilon)$ is a some scalar normalizing multiplier.
In their turn, from (2.36) and the statement (c) of Lemma 3.2,
it follows that
$$ \alpha_n(\varepsilon)=1+o(1),\qquad \varepsilon\to0.\tag3.2
$$ Due to  (3.1), (3.2) and (2.14), we obtain validity of
the statement c) of Theorem 1.1. Theorem is proved.
\enddemo

\heading{References}
\endheading

1. {\it Beale J.T.} Scattering Frequencies of Resonator,  Comm.
     Pure and Applied Math. 1973. V.~26. P.~549--564.

2. {\it Arsen'ev A.A.} On singularities of an analytic
continuation and resonance properties of the solution of the
scattering problem for the Helmholtz equation, Zh. Vychisl. Mat.
i Mat. Fiz.  1971. V.~12. P.~112--138 (in Russian). English
translation: Comput. Math. Math. Phys. 1972. V.~12. P.~139--173.

3. {\it Brown R.M., Hislop P.D., Martinez A.} Eigenvalues and
Resonances for Domains with Tubes: Neumann Boundary Conditions,
Jour. of Differential Equations. 1995. V.~115. P.~58--476.

4. {\it Gadyl'shin R.R.} On scattering by a cylinder wiyh a narrow
slit and walls of finite thickness, Teor. i Mat. Fiz. 1996.
V.~106. P.~24--43 (in Russian). English translation: Theor. Math.
Phys. 1996. V.~106. P.~19--34.

5. {\it Van Dyke M.D.} Perturbation Methods in Fluid Mechanics.
 Academic Press, New York. 1964.

6. {\it Nayfeh A.H.}  Perturbation Methods. John Wiley,
 New York. 1986.

7. {\it Il'in A.M.} Matching of Asymptotic  Expansions of
Solutions of Boundary-Value Problems. Nauka, Moscow. 1989 (in
Russian). English translation:  Amer. Math. Soc., Providence, RI.
1992.

8. {\it Gadyl'shin R.R.} On the scattering of $H$-polarized
electromagnetic field by an ideally conductive cylindrical body of
a trapping type,  Comptes Rendus Acad. Sci. Paris, Serie II~b.
2001. V.~329. P.~137--140.

9. {\it Sanchez-Palencia E.} Non-Homogeneous  Media  and Vibration
Theory. Springer-Verlag, New-York. 1980.

10. {\it Rayleigh.} The Theory of Helmholtz Resonator,  Proc. of
Royal Soc. London. 1916. V.~92. P.~265--275.

11.  {\it Arsen'ev A.A.} On the existence of resonances for
scattering in the case of boundary conditions of type II and III,
Zh. Vychisl. Mat. i Mat. Fiz. 1976. V.~16. P.~718--724 (in
Russian). English translation: Comput. Math. Math. Phys. 1976.
V.~12. P.~171--177.

12. {\it Gadyl'shin R.R.} On scattering frequencies of acoustic
resonator, Comptes Rendus Acad. Sci. Paris,  Serie I. 1993. V.316.
P.959-963.

13. {\it Gadyl'shin R.R.} Asymptotics of scattering frequencies
with small imaginary parts for acoustic resonator, Mathematical
Modelling and Numerical Analysis. 1994. V.~28. P.~761--780.

14. {\it Gadyl'shin R.R.} A two-dimensional analog of the
Helmholtz resonator with rigid walls. 1994. v.~30. P.~221--229 (in
Russian). English translation: Differ. Equations. V.~30.
P.~201--209.

15. {\it Lavrent'ev M.A., Shabat B.V.} Methods of the theory of
functions of complex variable. 4th ed. Nauka,  Moscow. 1973 (in
Russian). German translation of 3th ed.: VEB Deutscher Verlag
Wiss., Berlin. 1967.

16. {\it Gadyl'shin R.R.}  On the poles of an acoustic resonator,
Funktsional. Anal. i Prilozhen. 1993. V.~27(5). P.~3--16 (in
Russian). English translation: Functional Anal. Appl. 1993. V.~27.
P.~19--34.

17. {\it Gadyl'shin R.R.} On acoustic Helmholtz resonator and on
its
     electromagnetic ana\-lo\-gue,  J.  Math.  Phys.  1994. V.~35.
     P.~3464--3481.

\medskip
\noindent Bashkir State Pedagogical University, Ufa

\noindent e-mail: gadylshin\@bspu.ru
\enddocument
\end